\documentclass[11pt,superscriptaddress,showpacs,aps,preprint]{revtex4}
\usepackage[utf8]{inputenc}
\usepackage{cancel}
\usepackage{graphicx}
\begin{document}
\title{No radiative corrections to the Carroll-Field-Jackiw term beyond one-loop order}

\author{L. C. T. Brito}
\email{lcbrito@ufla.br}
\affiliation{Departamento de F\'{i}sica, Universidade Federal de Lavras, Caixa Postal 3037,
37200-000, Lavras, MG, Brasil}

\author{J. C. C. Felipe}
\email{jean.cfelipe@ufvjm.edu.br}
\affiliation{Instituto de Engenharia, Ci\^{e}ncia e Tecnologia, Universidade Federal dos Vales do Jequitinhonha e Mucuri, Avenida Um, nº 4050,  39447-814, Cidade Universitária,  Jana\'{u}ba,  MG, Brazil}

\author{A. Yu. Petrov}
\email{petrov@fisica.ufpb.br}
\affiliation{Departamento de F\'{i}sica, Universidade Federal da Para\'{i}ba, Caixa Postal 5008,
58051-970, Jo\~{a}o Pessoa, PB, Brazil}

\author{A. P. Ba\^eta Scarpelli}
\email{scarpelli@cefetmg.br}
\affiliation{Centro Federal de Educa\c{c}\~ao Tecnol\'ogica - MG \\
Avenida Amazonas, 7675, 30510-000,  Nova Gameleira - Belo Horizonte, MG, Brazil}


%%%%%%%%%
\begin{abstract}
We demonstrate explicitly the absence of the quantum corrections to the Carroll-Field-Jackiw (CFJ) term beyond one-loop  within the Lorentz-breaking CPT-odd extension of QED. The proof holds  within two prescriptions of quantum calculations, with the axial vector in the fermion sector {}treated either as  a perturbation  or as a contribution in the exact propagator of the fermion field. 
\end{abstract}
%%%%%%%%%

\maketitle

\section{Introduction}

 The Carroll-Field-Jackiw (CFJ) term ${\cal L}_{CFJ}=\epsilon^{\alpha\beta\gamma\delta}k_{\alpha}A_{\beta}\partial_{\gamma}A_{\delta}$, originally introduced in \cite{CFJ}, is certainly the most known and studied example of a Lorentz-breaking term.  In \cite{Jackiw:1999yp}, this term, for the first time, has been shown to arise  as a one-loop quantum correction in an appropriate Lorentz-breaking extension of QED, and turned out to be finite. Further, a number of results related to this term, with special attention to its  quantum generation, have been obtained. Besides being finite, the most important feature of this term is the fact that it is ambiguous in the one-loop approximation. From the formal viewpoint, its ambiguity is related with the fact that, actually, the Feynman diagrams contributing to this term are superficially divergent, so that one faces an undetermined expression like $\infty-\infty$. Further, various results for this term have been obtained within different calculation schemes (for a detailed discussion, see f.e. \cite{ourLV,perez} and references therein). This ambiguity was discussed from the physical viewpoint by Jackiw  \cite{JackAmb}, who pointed out that it is  an example of a situation where   radiative corrections are finite but still have to be fixed by experiments. In particular, this ambiguity  comes from ``triangle'' diagrams similar to those related with the Adler-Bell-Jackiw (ABJ) anomaly in QED \cite{ABJ}. In this paper, we argue that the one-loop result for the CFJ term is actually the unique possible, i.e., that no higher-loop CFJ contributions arise. Despite  similarities with calculation of the axial anomaly, it is evident that we cannot conclude from the  Adler-Bardeen theorem  that the CFJ term induced at one-loop order is exact ( see discussions in \cite{Adler:1969er,Adler:2004qt}).  Actually, we will reach this conclusion using the same  method  used  by Coleman and Hill  to prove that there is no radiative corrections to the Chern-Simons term beyond one-loop order in QED in $(2+1)$ dimensions (Coleman-Hill theorem) \cite{Coleman:1985zi}.  As we will see, this method is independent of the existence of anomalies and, therefore, is not based on the  arguments used to prove the Adler-Bardeen theorem \cite{Adler:1969er}.  It is important to mention that all known studies of the CFJ term have been performed at the one-loop order. Moreover, up to now, there is no examples of higher-loop calculations in Lorentz-breaking theories.  In this contribution, we take the first step towards this study. We will point out  that, as a  consequence of the Ward identities for the vectorial vertex, there is a non-renormalization theorem which  guarantees that beyond one loop any  Feynman diagram of the photon self-energy  in QED -- including the Lorentz-violating extension in  \cite{Jackiw:1999yp}  -- will be of second order in the external momentum; so there is no CFJ term induced from higher loops. 
 
 The paper is organized as follows: in section II, we review aspects related to the quantum induction of the CFJ term; in section III, we highlight the general property of the vertex functions with $n$ external photons, $\Gamma^{n}_{\mu_1,\dots,\mu_n}(p_1,\dots,p_n) $, namely, that it is at least quadratic in the external momentum for $n>2$;  in the section IV, we  apply the results of  section III  to the photon self-energy  in the Lorentz-violating electrodynamics proposed in Ref. \cite{Jackiw:1999yp}, making clear through examples why  it is impossible to induce the CFJ term beyond one-loop order; finally, in the section V, we formulate our conclusions.

\section{Radiative corrections and the CFJ term}
\begin{figure}[!htb]
\centering
\includegraphics[height=1.8 cm, width=10.0 cm] {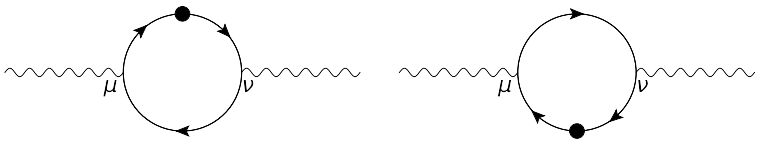}
\caption{ Two-point function of the gauge field.}
\label{fig1}
\end{figure}

Our starting point is the simplest Lorentz-breaking extension of QED  introduced already in  \cite{Jackiw:1999yp}:
\begin{equation}\label{model}
\mathcal{L}=-\frac{1}{4}F_{\mu\nu}F^{\mu\nu}+i\bar{\psi}\cancel{\partial}\psi-m\bar{\psi}\psi-e\bar{\psi}\gamma^{\mu}A_{\mu}\psi-\bar{\psi}\cancel{b}\gamma_{5}\psi,    
\end{equation}
in which $b_{\mu}$ is a constant  axial four-vector. Actually, it is easy to see that none of other minimal Lorentz-breaking extensions of QED \cite{KosPic} can yield the CFJ term in the first order in the Lorentz-breaking parameter, which is consistent with \cite{ouranom}. It is well known that, in the theory described by (\ref{model}), the  CFJ term 
\begin{equation}\label{cfj}
\Delta \mathcal{L}_{CFJ} = \frac{1}{2}c_{\mu}\epsilon^{\mu\alpha\beta\nu}F_{\alpha\beta}A_{\nu}
\end{equation}
can be induced  at one-loop order by radiative corrections arising from Lorentz and CPT violation in the  fermionic sector \cite{Jackiw:1999yp}. As it is well known (see f.e. \cite{Jackiw:1999yp}), when the axial term is treated as a perturbation, the Feynman diagrams generating the term (\ref{cfj}) looks like the triangle diagrams generating the ABJ anomaly in QED (see Fig. \ref{fig1}). 
In this section, in order to review some relevant points related  to this paper, we discuss in a bit more detail the similarities between  the diagrams in Fig. \ref{fig1} and the calculations leading to  the axial anomaly. 

\begin{figure}[!htb]
\centering
\includegraphics[height=8.0cm, width=10.0cm] {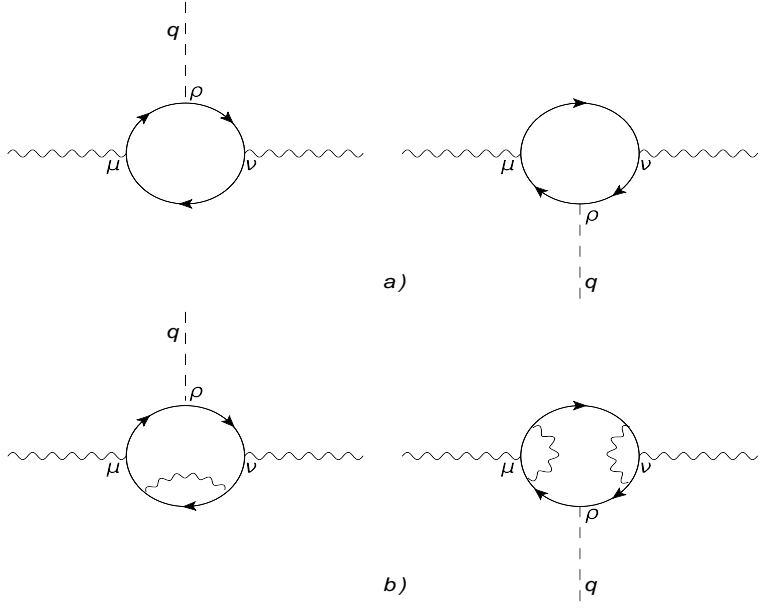}
\caption{ Contributions to the three-point spinor-vector function.}
\label{fig2}
\end{figure}

\begin{figure}[!htb]
\centering
\includegraphics[height=6.0cm, width=11.0cm] {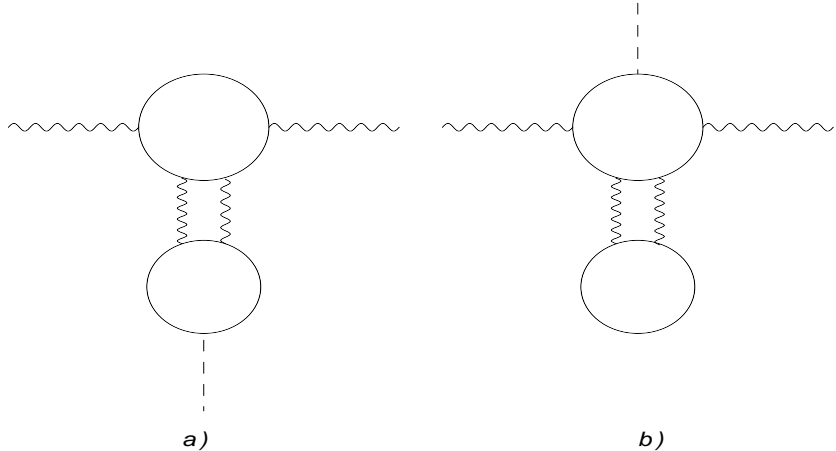}
\caption{Three-point diagrams which contribute (a) and do not contribute (b) to an anomalous amplitude.}
\label{fig-exemple}
\end{figure}

From the perturbative point of view, the existence of an anomaly in QED is a consequence of   ambiguities  arising from arbitrary shifts in the integration variable within quantum corrections \cite{Weinberg:1996kr,Bertlmann:1996xk}. This is the case for the triangle diagrams shown in Fig. \ref{fig2}a, which have a linear divergence from which the axial anomaly arises (the dashed line represents  an external background vector  field contracted to an axial vertex $\gamma^{\mu}\gamma_{5}$).  The Adler-Bardeen theorem \cite{Adler:1969er} proves that, at any order of the perturbative expansion, the  anomaly can arise only from diagrams containing the triangle subdiagrams given by the Fig. \ref{fig2}a. It follows from  the fact  that a fermion loop with one external axial field  and $n>2$  external photons will be at most logarithmically divergent, so that there is no ambiguity associated with shifts in the internal momenta. In this sense, for example, the diagram depicted in Fig. \ref{fig-exemple}a may contribute to an anomalous amplitude, while the diagram depicted at Fig. \ref{fig-exemple}b will not. This is because in the last case the external axial field (dashed line) appears  in a fermion loop involving four vertices, which is evidently convergent. From the same arguments, it is possible to conclude that no loop corrections to the triangle diagrams, such as the diagrams shown in Fig. \ref{fig2}b, will contribute to the anomaly. Besides, they satisfy the usual axial-vector Ward identity \cite{Adler:1969er}.

\begin{figure}[!htb]
\centering
\includegraphics[height=4.0cm, width=10.0cm] {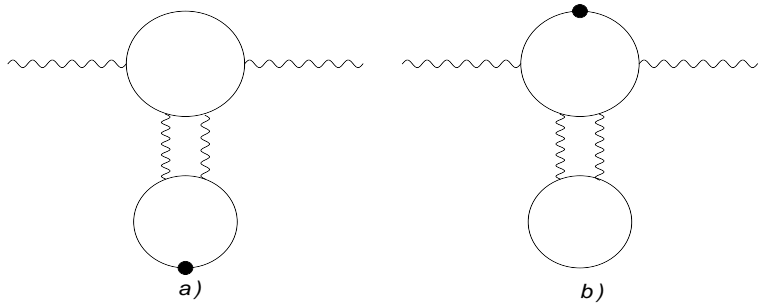}
\caption{Ambiguous and non-ambiguous contributions to the two-point function of the gauge field.}
\label{fig-exemple-2}
\end{figure}

Finally, we must formulate possible conclusions  about the induction of the CFJ term. The first one is obvious: if the axial vector $b_{\mu}$ is treated as a external field,  diagrams of the structure given at Fig. \ref{fig-exemple-2} will have the same properties of the diagrams given at Fig. \ref{fig-exemple} (with zero external momentum)   concerning shifts in internal momentum: those diagrams have ambiguities associated with integrals that are linearly divergent. Therefore, diagram \ref{fig-exemple-2}a  can in principle have an ambiguity of the same type occurring in the diagrams in Fig. \ref{fig1}, while the diagram in Fig. \ref{fig-exemple-2}b does not. The second conclusion, which is more important for our purpose, is that  diagrams  such  as those ones shown at Fig.  
\ref{fig-exemple-2} can both yield a  CFJ term in principle,  but without ambiguity coming from  the diagram in Fig. \ref{fig-exemple-2}b  at all. So, the only information  given by the Adler-Bardeen theorem in this context  is that an {\it ambiguous}  CFJ  term may arise only from diagrams which have diagrams like the ones given at Fig. \ref{fig1}a as sub-diagrams. Actually, we will show in the next sections that  there is no possibility that a CFJ term will be induced beyond one-loop order.  As we will see, this result still holds if the axial term in the Lagrangian (\ref{model}) is treated non-perturbatively \cite{perez}.
 
\section{The Coleman-Hill method}
\begin{figure}[!htb]
\centering
\includegraphics[height=2.0 cm, width=6.0 cm] {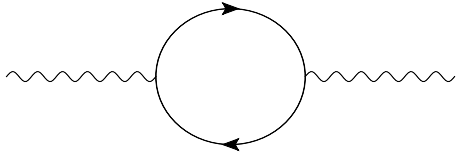}
\caption{Two-point function of the gauge field in $QED_3$.}
\label{QED3}
\end{figure}

In $(2+1)$ dimensions, it is possible to show that, beyond one-loop order,  all radiative corrections to the abelian Chern-Simons term \cite{Deser:1981wh}, 
\begin{equation}\label{chern}
\mathcal{L}_{CS} = \frac{\mu}{4} \epsilon^{\mu\nu\rho}F_{\mu\nu}A_{\rho},
\end{equation}
vanish identically. The parameter $\mu$ in the Lagrangian density (\ref{chern}) is the ``topological mass'' of the gauge field $A_{\mu}$. In particular, the only diagram giving (\ref{chern}) is the one-loop contribution to the photon self-energy of  ${\rm QED}_{3}$, shown in Fig. \ref{QED3}. This is the conclusion of the Coleman-Hill theorem \cite{Coleman:1985zi}. The result holds in a large class of theories and can be generalized to the non-Abelian case \cite{Brandt:2000yk,Brandt:2000qp}. 
In this section we will review the method used to prove the Coleman-Hill theorem. The essential feature of the method which is interesting for our purpose is that it is independent of the space-time dimension, and so can be applied directly to draw conclusions about the induction of the  CFJ term (\ref{cfj}) by radiative corrections. Indeed, it is possible since the terms (\ref{cfj})  and  (\ref{chern}) are both linear in the momentum.

Actually, the method  has very solid foundations, since it is based on the  Ward identities  for the vertex function $\Gamma^{(n)}_{\mu_{1}\cdots\mu_{n}}(p_{1},\cdots,p_{n})$ with $n$ external photon lines,  looking like
 \begin{equation}\label{ward}
 p^{\mu_{1}}_{i}\Gamma^{(n)}_{\mu_{1},\cdots,\mu_{n}}(p_{1},\cdots,p_{n})=0,
 \end{equation}
which represents the transversality of all quantum corrections to any $n$-point function of the gauge field. Here, the indices in the momenta $p_{i}$ of the external photons take values $i=1,2,\cdots,n$. 

The main premise we assume is that the vertex functions $\Gamma^{(n)}$ are  analytic in the limit of zero external momenta. In order to avoid  infrared singularities in this limit, we maintain $m \neq 0$ in the fermion sector of the Lagrangian density (\ref{model}).  At the same time, it is well known that singularities  arising from photon propagators inside a arbitrary Feynman diagram  do not cause  infrared problems. Of course, we also assume that  the regularization used to handle divergent integrals is gauge invariant.

Next, we will follow  the methodology used in \cite{Coleman:1985zi}  and  show that for $n>2$, the vertex functions $\Gamma^{(n)}_{\mu_{1},\cdots,\mu_{n}}(p_{1},\cdots,p_{n})$ are $\mathcal{O}(p_{1}p_{2})$, where $p_{1}$ and $p_{2}$ are independent  external momenta.  Indeed, differentiating (\ref{ward}) with respect to $p_{1}$  and then, setting $p_1$ to zero we find
\begin{equation}
\Gamma^{(n)}_{\mu_{1},\cdots,\mu_{n}}(0,p_{2},\cdots,p_{n})= 0.
\end{equation}
Therefore,  it follows from the Taylor expansion that
\begin{equation}\label{condP1}
\Gamma^{(n)}_{\mu_{1},\cdots,\mu_{n}}(p_{1},p_{2},\cdots,p_{n})=\mathcal{O}(p_1).
\end{equation}
 Since we  have chosen $p_{1}$ and $p_{2}$  to be independent variables,  which is possible for $n>2$,  we can apply these arguments as well to $p_2$   and find
\begin{equation}\label{condP2}
\Gamma^{(n)}_{\mu_{1},\cdots,\mu_{n}}(p_{1},p_{2},\cdots,p_{n})= \mathcal{O}(p_2).
\end{equation}
As the conditions (\ref{condP1}) and (\ref{condP2}) are independent and must be simultaneously satisfied, we conclude that
\begin{equation}\label{condP1P2}
\Gamma^{(n)}_{\mu_{1},\cdots,\mu_{n}}(p_{1},p_{2},\cdots,p_{n})= \mathcal{O}(p_1 p_2).
\end{equation}
This is the key relation we need to prove the analogue of the Coleman-Hill theorem for the Lorentz-violating theory given by the Lagrangian density (\ref{model}).

\section{Higher loop contributions in the photon self-energy}

Now we will present examples that make clear from the result given in ({\ref{condP1P2})  that the CFJ term arising from the diagrams in Fig. \ref{fig1} is exact. It follows directly from the fact that any diagram in the photon self-energy can be expressed in terms of  diagrams contributing to the} vertex function $\Gamma^{(n)}_{\mu_{1},\cdots,\mu_{n}}(p_{1},p_{2},\cdots,p_{n})$ with $n>2$; therefore, when we sum all diagrams in a certain order of perturbation, the contribution is at least of the order $p_1 p_2$.  Without loss of generality, in what follows we will assume that $p_1$ and $p_2$ correspond to the external  lines in the photon self-energy. For simplicity, in this section we will consider that  the fermion propagator  is the exact propagator \cite{perez}
\begin{equation}\label{exact}
S_{b}(p)=\frac{i}{\cancel{p}-m-\cancel{b}\gamma_5}.
\end{equation}
So the diagrammatic representation  of the photon self-energy is the same occurring  in usual quantum electrodynamics  but with fermion lines given by ({\ref{exact}). At the end we will comment on the case that $\cancel{b}\gamma_5$ is treated as a perturbation; of course, it does not change the general result.
 \begin{figure}[!htb]
\centering
\includegraphics[height=7.0cm, width=12.0cm] {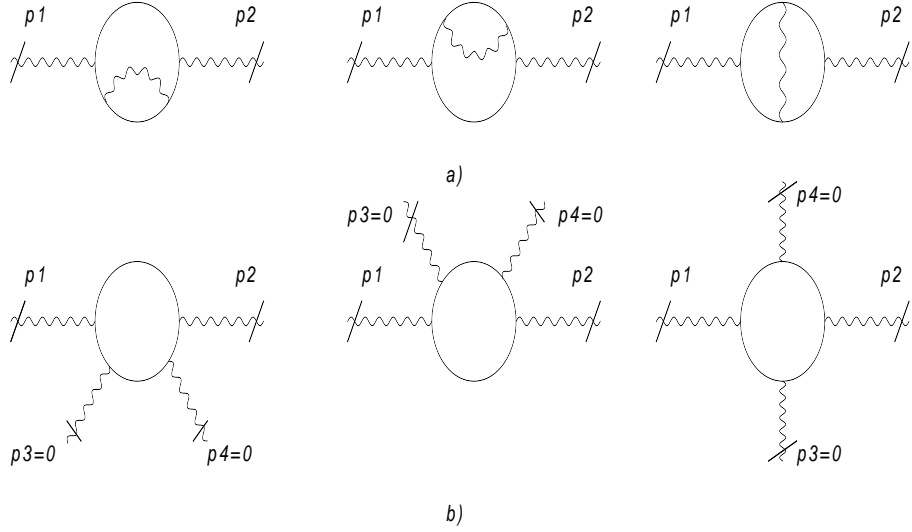}
\caption{ Two-loops diagrams in (a) can be written in terms of the diagrams in figura (b) by cutting the internal lines of the photon propagator.}
\label{2loops-1}
\end{figure} 
Let us start with two-loop diagrams in the photon self-energy, given  in Fig. \ref{2loops-1}a. We can always ``cut'' the internal lines of those diagrams and express them in terms of  the diagrams in Fig. \ref{2loops-1}b. So, at two-loop order, the photon self-energy can be expressed in terms of the sum of the diagrams in Fig. \ref{2loops-1}b, that is, of the  vertex function  $\Gamma^{(4)}_{\mu,\nu,\rho,\sigma}(p_{1},p_{2},p_{3},p_{4})$ with $p_{3}=p_{4}=0$. So, at the two-loop order, the self-energy is at least quadratic in momenta being proportional to $p_1 p_2$ and, therefore, does not induce a CFJ term at this order.

Next, we will consider more general diagrams like the ones shown in Fig. \ref{highloops}. It is evident that by cutting the internal photon lines in those diagrams we express them in terms of a vertex function $\Gamma^{(n)}_{\mu_{1}\cdots\mu_{n}}(p_{1},\cdots,p_{n})$ containing the external momenta $p_1$ and $p_2$ and with the momenta in the others lines vanishing. Thus, from (\ref{condP1P2}) we conclude that  both diagrams are at least of order $p_1 p_2$. Again, the CFJ term is not induced from this class of diagrams.

\begin{figure}[!htb]
\centering
\includegraphics[height=4.0cm, width=11.0cm] {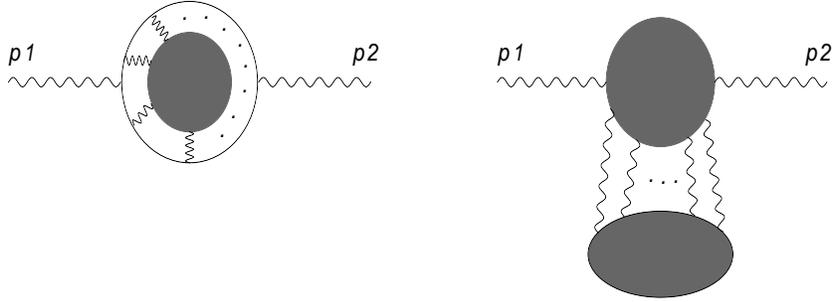}
\caption{ High-loops diagrams which are quadratic in the external momenta..}
\label{highloops}
\end{figure}

Following the same arguments presented above, we conclude that the CFJ term is induced in the theory defined by the Lagrangian density (\ref{model}) only by the diagram shown in Fig. \ref{QED3}.  We can always express any other diagram in terms of the $n$-point  vertex function of the photon by cutting internal lines and putting zero momenta in these lines, such that the total contribution will be at least of second order in the external momentum. In fact, these results are valid for a specific diagram. We can verify that individual diagrams give rise to  a CFJ term. But, as a consequence of gauge invariance, terms coming from different  diagrams in a certain order of perturbation must cancel like was explicitly verified in the (2+1) case \cite{chen,Kao:1984ft}. Note that the results presented in this section are immediately applicable in a perturbative approach. Indeed,  diagrams with insertions like the ones shown in figures \ref{fig1} and \ref{fig-exemple-2} are easily obtained by Taylor expansion of the exact propagator (\ref{exact}).

Finally, it is interesting to mention  situations where the hypothesis of the  Coleman-Hill theorem is not satisfied.  One interesting example is when the gauge symmetry is spontaneously broken. In fact, in the (2+1) dimensions the abelian Chern-Simons term (\ref{chern}) is induced at the one-loop order as a consequence of the spontaneous breaking of the $U(1)$ gauge symmetry \cite{Khare:1994yv}. The corresponding result for the CFJ {\bf term} in  Lorentz-violating theories was actually found in Ref. \cite{Brito:2013npa}. Of course, in this case we also may have a non-vanishing result at higher-loop orders. Other relevant case  is when the theory involves massless particles. It was explicitly verified in this case that in (2+1) dimensions there is a finite renormalization  of the Chern-Simons in two loops  \cite{chen,semenoff,spiridonov}.  It is possible that a well-defined and finite  CFJ term can be induced at {\bf the} two-loop order in a Lorentz-violating theory with massless scalars, as occurs with the Chern-Simons term in  (2+1) dimensions.

\section{Conclusion}
 Using the properties of the Feynman diagrams, we proved the absence of a CFJ-like correction beyond one-loop order in the quantum electrodynamics modified by an axial vector in the fermion sector. The essence of our proof  follows the method used in \cite{Coleman:1985zi}} to prove the Coleman-Hill theorem in 2+1 dimensions. First, we observe that any two-loop contribution to the two-point function of the gauge field can be  obtained as a contraction of external legs in some one-loop four-point function of the gauge field. Since, as we already noted, each such four-point function, by gauge invariance reasons, is at least quadratic in external momenta, we note that, after the contraction of some gauge fields into the propagators, the corresponding two-point function of the gauge field will also be at least quadratic in the external momentum and, thus, will not contribute to the CFJ term. Clearly, the same argumentation can be applied for higher-point contributions to the one-loop effective action of the gauge field, which also are at least quadratic in external momenta by gauge symmetry reasons. After performing similar contractions of some external gauge legs to the propagators, we arrive at higher-loop results which are also quadratic in the external momentum and, hence, do not contribute to the CFJ term. Effectively, we proved that the unique possible correction to the CFJ term is of one-loop order.   Also, it is interesting to note that in theories containing massless particles or spontaneous breaking of gauge symmetries the hypothesis on which the proof is based does not hold, thus it is possible that a well defined and  finite  CFJ term can be induced at higher orders. The results presented in this paper can be applied to non-Abelian theories as well using similar argumentation and, moreover, may have some interest in application of the Lorentz-violating field theories to  condensed matter systems \cite{Casana:2012ki}. Another possible extension of this paper could consist in generalizing our proof for other minimal Lorentz-breaking extensions of QED presented in \cite{KosPic}. We plan to perform these studies in forthcoming papers.

 {\bf Acknowledgments.} The work by A. Yu. P. has been partially supported by the
CNPq project No. 301562/2019-9. APBS thanks CNPq for financial support.

\end{document}